\documentclass[notitlepage, twoside,english,nofootinbib,preprintnumbers,aps,prd,11pt,showpacs, superscriptaddress, groupedaddress]{revtex4-2}
\usepackage[utf8]{inputenc}
\usepackage[english]{babel}
\usepackage[T1]{fontenc}
\usepackage{lmodern}
\usepackage[intlimits]{amsmath}
\usepackage{amssymb}
\usepackage{amsfonts}
\usepackage{mathrsfs}
\usepackage{physics}
\usepackage{tensor}
\usepackage{enumerate}
\usepackage{mathtools}
\usepackage{csquotes}
\usepackage{scalerel,stackengine}
\usepackage{hyperref}
\hypersetup{
   breaklinks=true,
   colorlinks=true,
   urlcolor=blue
}
\usepackage{xurl}
\usepackage{cancel}

\renewcommand{\t}{\tensor}

\renewcommand{\r}{\frac}

\newcommand{\m}[1]{\begin{bmatrix}#1\end{bmatrix}}

\newcommand{\mc}[1]{\mathcal{#1}}

\newcommand{\mr}[1]{\mathrm{#1}}
\newcommand{\mf}[1]{\mathfrak{#1}}

\newcommand{\ar}{\rightarrow}

\newcommand{\f}{\qty}

\newcommand{\ti}[1]{\tilde{#1}}
\newcommand{\pd}{\partial}
\newcommand{\mri}[1]{\mathring{#1}}

\DeclareMathOperator{\diag}{diag}

\newcommand{\scri}{\mathscr{I}}
\newcommand{\lie}{\mathcal{L}}
\newcommand{\lieg}{\mathcal{L}^{\Gamma}}
\newcommand{\liegp}{\mathcal{L}^{\Gamma'}}

\newcommand{\ipr}{\mathbin{\lrcorner}}
\newcommand{\hod}{{\star}}

\begin{document}

\title{Symplectic charges in the Yang-Mills theory of the normal conformal Cartan connection: Applications to gravity}

\author{Adam Bac}
\email[]{ab417503@okwf.fuw.edu.pl}
\author{Wojciech Kamiński}
\email[]{Wojciech.Kaminski@fuw.edu.pl}
\author{Jerzy Lewandowski}
\email[]{Jerzy.Lewandowski@fuw.edu.pl}
\author{Michalina Broda}
\email[]{ma.broda@student.uw.edu.pl}
\affiliation{Faculty of Physics, University of Warsaw, Pasteura 5, 02-093 Warsaw, Poland}

\date{\today}

\begin{abstract}
It is known that a source-free Yang-Mills theory with the normal conformal Cartan connection used as the gauge potential gives rise to equations of motion equivalent to the vanishing of the Bach tensor. We investigate the conformally invariant presymplectic potential current obtained from this theory and find that on the solutions to the Einstein field equations, it can be decomposed into a topological term derived from the Euler density and a part proportional to the potential of the standard Einstein-Hilbert Lagrangian. The pullback of our potential to the asymptotic boundary of an asymptotically de Sitter spacetime turns out to coincide with the current obtained from the holographically renormalized gravitational action. This provides an alternative derivation of a symplectic structure on scri without resorting to holographic techniques. We also calculate our current at the null infinity of an asymptotically flat spacetime and in particular show that it vanishes for variations induced by the Bondi-Metzner-Sachs group of asymptotic symmetries. In addition, we calculate the Noether currents and charges corresponding to gauge transformations and diffeomorphisms.
\end{abstract}

\maketitle

\section{Introduction}
The interest in the conformal completions of Einstein spacetimes \cite{penrose_asymptotic_1963,geroch_asymptotic_1977,compere_setting_2008,ashtekar_asymptotics_2014,compere_lbms4_2019,compere_lbms4_2020,kolanowski_hamiltonian_2021,senovilla_gravitational_2022} stems from their applications in the analysis of the asymptotics of solutions, as well as the holographic theory. It leads to considerations that use both Einstein's equations and conformal geometry simultaneously.
This is not straightforward since Einstein's equations are not conformally invariant. In fact, they become singular at the conformal boundary, which makes the analysis of the equations complicated. Moreover, natural structures such as the symplectic potential and Noether charges cannot be defined at the boundary in a straightforward way. For example, one needs to perform a renormalization procedure to obtain a symplectic potential at the conformal boundary \cite{deharo_holographic_2001,compere_setting_2008}. The regularization is defined in a specific gauge, the so-called Fefferman-Starobinski coordinate system \cite{compere_setting_2008, compere_lbms4_2019}. It is interesting to try to find another, more natural way of obtaining the pullback of the potential. 

In four-dimensional spacetimes, a conformally invariant condition for satisfying vacuum Einstein's equations with a (possibly vanishing) cosmological constant $\Lambda$ is the vanishing of the Bach tensor \cite{kozameh_conformal_1985}. This way, the space of Einstein metrics is naturally immersed in the space of Bach flat metrics. Bach's equations were used to study the structure and stability of the conformal boundary in Einstein spacetimes \cite{anderson_existence_2005a,anderson_asymptotically_2005,kaminski_wellposedness_2022a}. They share many properties with Einstein's field equations. In particular, they form a well-posed evolution system \cite{guenther_ueber_1975,kaminski_wellposedness_2022a} and are obtained from a Lagrangian \cite{bach_zur_1921}. The success of the application of these equations to general relativity leads to a natural question of whether the symplectic potential or Noether currents of this auxiliary theory can be of some use in the Einstein theory.

In this paper, we propose a possible way of exploiting the relation between the spaces of Einstein and Bach flat metrics by pulling back into the former the geometrical conformally invariant structure with which the latter is equipped. We construct a conformally invariant presymplectic potential current for the Bach theory and show that its pullback to the space of Einstein metric tensors with nonvanishing cosmological constant differs from the presymplectic potential current for the Einstein theory by trivial terms. The advantage of this approach is that due to the conformal invariance, our current is automatically finite at the conformal boundary (scri) of asymptotically (anti-)de Sitter spacetimes. This way, one can obtain currents that are well defined at all spacetime points simultaneously, both in the bulk and on scri. We show, using the Fefferman-Graham coordinates, that the pullback of this current to scri of an asymptotically de Sitter spacetime, coincides with the symplectic potential current obtained by the method of holographic renormalization \cite{compere_setting_2008, compere_lbms4_2019}. For the sake of completeness, we also consider the case of an asymptotically flat spacetime and calculate the new symplectic potential on its null boundary. However, since in the case of vanishing cosmological constant the potential only consists of a trivial, topological term, the result lacks a clear physical meaning and in fact vanishes for variations generated by the diffeomorphisms belonging to the Bondi-Metzner-Sachs (BMS) group of asymptotic symmetries of an asymptotically flat spacetime. 

To achieve this, we will use two different Lagrangians for the Bach theory, both of which have interesting but different properties. They differ by a topological term (the Euler-Gauss-Bonnet term \cite{cherubini_second_2002}), and thus their symplectic currents are related.
The first Lagrangian is conformally invariant, which we ensure by using the normal conformal Cartan connection \cite{cartan_espaces_1923,kobayashi_transformation_1972,korzynski_normal_2003,curry_introduction_2018,merkulov_conformally_1984,friedrich_twistor_1977} (closely related to the local twistor connection \cite{penrose_spinors_1986,merkulov_conformally_1984,friedrich_twistor_1977} as well as to the tractor calculus \cite{curry_introduction_2018,herfray_einstein_2022,herfray_asymptotic_2020}) as the gauge potential in the (also conformally invariant) Yang-Mills Lagrangian.
The corresponding symplectic potential current defined in terms of the Cartan connection is conformally invariant. The second Lagrangian is not conformally invariant; however, its symplectic current restricted to the space of solutions of Einstein field equations with cosmological constant is proportional to the symplectic potential of general relativity.
The pullback of the Yang-Mills symplectic potential current to the space of Einstein metric tensors with a nonvanishing cosmological constant becomes a linear combination of symplectic potential currents of the gravitational Lagrangian and the Euler Lagrangian, respectively.    

\subsection{Notation and conventions}

We consider a four-dimensional manifold equipped with a (pseudo-)Riemannian metric. In Secs. \ref{sec-cnc} and \ref{sec-cym} the metric can have any nondegenerate signature, while in Sec. \ref{sec-boundary} we restrict ourselves to spacetimes with the Lorentzian signature $\diag(-,+,+,+)$.

To express fields on the manifold we will use both orthonormal tetrads, as well as holonomic tetrads corresponding to a coordinate system. In the first case, we shall use the indices $a,b,c,\ldots = 0,1,2,3$ to enumerate the elements of the tangent and cotangent tetrads, components of tensors with respect to those tetrads, and tetrad connection one-forms. On the other hand, to express tensors and the metric connection in terms of a coordinate system we will use greek letters $\alpha,\beta,\gamma,\ldots = 0,1,2,3$. The normal conformal Cartan connection and its curvature have values in a 6 by 6 matrix algebra -- we will use the uppercase indices $I,J,K,\ldots = 0,\ldots,5$ to identify their components. In Sec. \ref{sec-boundary} we also use indices $i,j,k,\ldots = 1,2,3$ and $A,B,C,\ldots = 2,3$ to identify subsets of the Fefferman-Graham and Bondi-Sachs coordinates, in a way which is explained in more details in the corresponding subsections.

\section{\label{sec-cnc}Normal conformal Cartan connection}

\subsection{Definition}
We will now provide a working definition of the normal conformal Cartan (NCC) connection. A more elegant geometric definition of this structure can be found in  \cite{kobayashi_transformation_1972} and the relation between the two formulations is explained in \cite{korzynski_normal_2003}. It is also worth mentioning that Penrose's notion of the local twistor connection 
\cite{penrose_spinors_1986,merkulov_conformally_1984,friedrich_twistor_1977} is simply the spinorial version of the NCC connection. 

We start by considering a four-dimensional manifold $M$ endowed with a spacetime metric tensor 
\begin{equation}\label{coframe}
 g = \eta_{ab}\theta^{a}\otimes \theta^{b},
\end{equation}
where $\eta_{ab}$ is a constant, nondegenerate, symmetric matrix of signature $(p,q)$, and $(\theta^0,\ldots,\theta^3)$ is a locally defined cotangent frame dual to a tangent frame  $(e_0,\ldots,e_3)$. The matrix $\eta_{ab}$ and its inverse  $\eta^{ab}$ will be used to raise and lower the tetrad indices $a,b,c,\ldots$. The choice of a coframe defines a volume form
\begin{equation}
\begin{split}
    \mr{Vol} &\coloneqq \r{1}{4!}\sqrt{\abs{\det\eta}} \varepsilon_{abcd} \theta^a \wedge \theta^b \wedge \theta^c \wedge \theta^d \\
    &= \sqrt{\abs{\det\eta}} \theta^0 \wedge \theta^1 \wedge \theta^2 \wedge \theta^3,
\end{split}
\end{equation}
whose components we will denote by $\epsilon_{abcd} = \sqrt{\abs{\det\eta}}\varepsilon_{abcd}$.

The CNC connection corresponding to a given choice of tetrad $\theta^a$ is defined as a matrix of one-forms, 
\begin{equation}\label{CCNC}
    A = \m{0 & \theta_b & 0 \\ P^a & \t{\Gamma}{^a_b} & \theta^a \\ 0 & P_{b} & 0 },
\end{equation}
where $\t{\Gamma}{^a_b}$ are the tetrad connection one-forms defined by the properties 
\begin{equation}\label{Gamma}
    \dd{\theta^a} + \t{\Gamma}{^a_b}\wedge\theta^b = 0,
    \quad \Gamma_{ab} = -\Gamma_{ba},
\end{equation}
while the one-forms $P_a=P_{ab}\theta^b$ are defined by minus the Schouten tensor, that is
\begin{equation}
    P_a := \f(\r{1}{12}R\eta_{ab} - \r{1}{2}R_{ab})\theta^b,
\end{equation}
where $R_{ab}$ and $R$ are the Ricci tensor and Ricci scalar, respectively. In this notation, the Riemann tensor of $g$ can be represented as a two-form
\begin{equation}
\t{\mc{R}}{^a_b} \coloneqq \frac{1}{2}R^a{}_{bcd}\theta^c\wedge \theta^d = \dd{\Gamma^a{}_b} + \Gamma^a{}_c\wedge \Gamma^c{}_b.
\end{equation}

\subsection{Conformal rescalings  and Lorenz transformations}\label{sec-transformations}
The special property of the CNC connection is the way it transforms upon rescalings \cite{korzynski_normal_2003}
\begin{equation}
g'=f^2 g, \quad \theta'^a = f\theta^a, \quad f\in C^\infty(M),
\end{equation}
namely
\begin{equation}\label{homegah}
A'= h^{-1}Ah + h^{-1}\dd{h},
\end{equation}
with
\begin{equation}\label{h-conformal}
    h = \m{\frac{1}{f} & 0 & 0 \\[5pt] \frac{\t{f}{_,^a}}{f^2} & \delta^a{}_b & 0 \\[5pt] \frac{f_{,c}f_{,}{}^{c}}{2f^3} & \frac{f_{,b}}{f} & f },
\end{equation}
where 
\begin{equation}
f_{,a} \theta^a := \dd{f}.   
\end{equation}

Of course, for the local (pseudo)rotations, that is   
\begin{equation}
g'=g, \quad \theta'^a = \t{\Lambda}{^a_b}\theta^b,   
\end{equation}
where $\eta_{ab}\t{\Lambda}{^a_c}\t{\Lambda}{^b_d} = \eta_{cd}$, the transformation law (\ref{homegah}) still applies, with
\begin{equation}\label{h-lorentz}
    h = \m{1 & 0 & 0 \\ 0 & \Lambda^a{}_b & 0 \\ 0 & 0 & 1 }.
\end{equation}

Both the matrix (\ref{h-conformal}) corresponding to the conformal rescaling of the tetrad and (\ref{h-lorentz}) associated with a (pseudo)rotation represent elements of the group $SO(p+1,q+1)$ identified with the group $SO(Q)$ of 6 by 6 matrices $h$ which preserve the form $Q$,
\begin{equation}
    \t{h}{^I_K}\t{h}{^J_L}Q_{IJ} = Q_{KL},
\end{equation}
with
\begin{equation}\label{form-q}
    Q = \m{0 & 0 & -1 \\ 0 & \eta & 0 \\ -1 & 0 & 0}.
\end{equation}

\subsection{Curvature}
The curvature of the CNC connection is represented by a matrix obtained from $A$ by the standard formula
\begin{equation}
    F = \dd{A} + A \wedge A
\end{equation}
and a calculation yields
\begin{equation}\label{fc}
    F = \m{0 & 0 & 0 \\ DP^a & \t{C}{^a_b} & 0 \\ 0 & DP_b & 0 },
\end{equation}
where
\begin{equation}
DP^a = \dd{P^a} + \Gamma^a{}_b\wedge P^b, \ \ \ C^a{}_b = \frac{1}{2}C^a{}_{bcd}\theta^c\wedge\theta^d
\end{equation}
and $C^a{}_{bcd}$ stands for the Weyl tensor of $g$. The curvature satisfies the Bianchi identity
\begin{equation}
D_A F := \dd{F} + A\wedge F - F\wedge  A =0
\end{equation}
and the transformation law
\begin{equation}\label{hFh} 
F'= h^{-1}Fh 
\end{equation}
under a transformation of $A$ given by (\ref{homegah}) for any $h \in SO(Q)$. As a result, complex identities of Riemannian geometry satisfied by the Weyl tensor can be written in an elegant and graceful way. 

\subsection{Relation with the Bach tensor}
Another conformally invariant operation in four-dimensional geometry is the Hodge dual of a differential two-form. We can apply it to the curvature $F$ and find that \cite{merkulov_conformally_1984,korzynski_normal_2003}
\begin{equation}\label{D*F}
    D_A{\hod F}:= \dd{\hod F} + A\wedge \hod F - \hod F\wedge  A = \m{0&0&0 \\ B^{ac}{\hod \theta}_c & 0 & 0 \\ 0 & \t{B}{_b^c}{\hod \theta}_c & 0 },
\end{equation}
where $B_{ab}$ is the Bach tensor, 
\begin{equation}
B_{ab} = 2\nabla^c \nabla_{[b} P_{c]a} - 2P^{cd}C_{cadb}.     
\end{equation}
What is particularly important from the point of view of applications to Einstein's theory of gravity, is that 
\begin{equation}\label{B=0}
R_{ab}=\Lambda \eta_{ab}\ \  \implies\ \ B_{ab}=0
\end{equation}
and that the second equality above is conformally invariant. Hence, the Bach tensor is an obstacle to the metric tensor being conformally Einstein. Its vanishing is not a sufficient condition, though -- there are known examples of spacetimes with $B_{ab} = 0$ which are not conformally Einstein \cite{lewandowski_fefferman_1988,nurowski_nonvacuum_2001,korzynski_normal_2003}, found among the homogeneous Fefferman metric tensors. As we can now see, in terms of the CNC connection, this obstacle becomes $D{\hod F}$.

\section{\label{sec-cym}Cartan-Yang-Mills theory}
\subsection{Cartan-Yang-Mills Lagrangian}
Inspired by (\ref{D*F}) and (\ref{B=0}) we define on the space of $\eta$-orthonormal coframes $\theta^a$ (\ref{coframe}) a Lagrangian by inserting the CNC connection $A$ into the standard Yang-Mills Lagrangian,
\begin{equation}\label{cym-lagrangian}
    L_{\mr{CYM}}(\theta) = \r{1}{2}\t{F}{^I_J}\wedge \t{\hod F}{^J_I},
\end{equation}
and name it the Cartan-Yang-Mills Lagrangian. The coframes are defined locally on $M$; however, the Lagrangian is independent of the transformations (\ref{homegah}) -- hence, it is uniquely defined on the entire $M$. Moreover,  
\begin{equation}
 L_{\mr{CYM}}(\theta) =  L_{\mr{CYM}}(f\theta), \ \ \ f\in C^3(M);
\end{equation}
hence, the Lagrangian  is manifestly conformally invariant. As a matter of fact, it follows from (\ref{fc}) that
\begin{equation}\label{cym-lagrangian-weyl}
L_{\mr{CYM}}(\theta) = \r{1}{2}\t{F}{^I_J}\wedge \t{\hod F}{^J_I} = \r{1}{2}\t{C}{^a_b}\wedge \t{\hod C}{^b_a} = \frac{1}{4}C_{abcd}C^{abcd} \mr{Vol},
\end{equation}
which makes it clear that despite of $L_{\mr{CYM}}$ being defined as a function of $\theta^a$, it depends only on $g = \eta_{ab}\theta^a\theta^b$. The right-hand side (RHS) of (\ref{cym-lagrangian-weyl}) is encountered in literature. However, we want to take advantage of the particular properties of the CNC connection and the Yang-Mills Lagrangian.

\subsection{Variations and the field equations}
Varying the Lagrangian with respect to $\theta^a$ and ``integrating by parts,'' we obtain
\begin{equation}
 \delta L_{\mr{CYM}}(\theta) = \delta \t{A}{^I_J}\wedge D_A \t{\hod F}{^J_I} + \r{1}{2}\t{F}{^I_J}\wedge \t{(\delta \hod )F}{^J_I} + \dd{\f(\delta \t{A}{^I_J}\wedge {\hod \t{F}{^J_I}})}.
\end{equation}
The term involving $\delta \hod $ vanishes due to the fact that the left and right duals of the Weyl tensor coincide \cite{batista_weyl_2013}: 
\begin{equation}\label{weyl-dual}
\hod C^{a}{}_{b} = \frac{1}{2}\epsilon^a{}_b{}_c{}^dC^c{}_d,
\end{equation}
where $\epsilon_{abcd} = \sqrt{\abs{\det\eta}}\varepsilon_{abcd}$; hence, it is constant, independent of $\theta^a$.  Breaking down the first term, we obtain \cite{korzynski_normal_2003}
\begin{equation}\label{deltaL}
 \delta L_{\mr{CYM}}(\theta) = 2\delta \t{\theta}{^a}\wedge {B}_{ab}{\hod \theta}^b  + \dd{\f(\delta \t{A}{^I_J}\wedge \hod  \t{F}{^J_I})}.
\end{equation}
From the first term, we obtain the equations of the theory,
\begin{equation}\label{Bach}
B_{ab} = 0, 
\end{equation}
equivalent to 
\begin{equation}
D_A {\hod F} = 0.    
\end{equation}

\subsection{Symplectic current potential}
From the second, exact term in the RHS of (\ref{deltaL}) we obtain the symplectic current potential
\begin{equation}\label{ThetaCYM1}
\Theta_{\mr{CYM}}(\theta; \delta\theta) = \delta \t{A}{^I_J}\wedge \hod  \t{F}{^J_I}.    
\end{equation}
It is conformally invariant due to (\ref{homegah}) and (\ref{hFh}),
\begin{equation}\label{theta-conformal-invariance}
\begin{split}
\f(\delta \t{A}{^I_J}\wedge \hod  \t{F}{^J_I})(f \theta; f\delta \theta)
&= \f(\left(h^{-1}\delta Ah\right)^I{}_J\wedge \hod  \left(h^{-1}Fh\right)^J{}_I)(\theta; \delta\theta) \\
&= \f(\delta \t{A}{^I_J}\wedge \hod  \t{F}{^J_I})(\theta; \delta\theta).
\end{split}
\end{equation}
A short calculation, using the explicit form of the CNC connection and its curvature associated with a given orthonormal tetrad (\ref{CCNC}), (\ref{fc}), gives the detailed form
\begin{equation}\label{ThetaCYM2}
    \Theta_{\mr{CYM}}(\theta;\delta\theta) = 2\delta \theta^a\wedge \hod DP_a + \delta \t{\Gamma}{^a_b}\wedge\t{\hod C}{^b_a}.
\end{equation}

\subsection{\texorpdfstring{$\Theta_{\mr{CYM}}$}{Potential current} at Einstein metrics}

When $\theta^a$ defines an Einstein metric tensor, that is when 
\begin{equation}\label{Ee}
R_{ab}=\Lambda \eta_{ab},
\end{equation}
our very symplectic current potential (\ref{ThetaCYM2}) takes a simpler form,
\begin{equation}\label{ThetaCYM3}
\Theta_{\mr{CYM}}(\theta; \delta\theta) =  \delta \t{\Gamma}{^a_b}\wedge\t{\hod C}{^b_a},
\end{equation}
where the first term in (\ref{ThetaCYM2}) vanishes due to the fact that (\ref{Ee}) implies $P_a = -\r{\Lambda}{6}\theta_a$ and $D\theta_a = 0$.

Since the Einstein metric tensors satisfy the equations of the Cartan-Yang-Mills theory (as their Bach tensor vanishes), all currents obtained from the three-form $\Theta_{\mr{CYM}}$ also apply to them. Therefore a natural question arises: what is the relation between $\Theta_{\mr{CYM}}$ and the symplectic current potential $\Theta_{\mr{EH}}$ of the Einstein-Hilbert action extended to the space of the $\eta$-orthonormal coframes?    

To answer this question, we will decompose $L_{\mr{CYM}}$ into two parts
\begin{equation}\label{lcym-euler-decomposition}
    L_{\mr{CYM}} = \r{1}{4}\mc{E} + L_1,
\end{equation}
where the Euler term \cite{cherubini_second_2002}
\begin{equation}
    \mc{E}(\theta) \coloneqq \epsilon^{abcd}\mc{R}_{ab}\wedge \mc{R}_{cd}
\end{equation}
is a topological Lagrangian whose variation only yields the boundary term:
\begin{equation}\label{euler-variation}
\begin{split}
    \delta\mc{E}(\theta)
    &= 2\epsilon^{abcd}\delta \mc{R}_{ab}\wedge \mc{R}_{cd} \\
    &= 2\epsilon^{abcd}D\f(\delta \Gamma_{ab})\wedge \mc{R}_{cd} \\
    &= 2\epsilon^{abcd}\dd{\f(\delta \Gamma_{ab}\wedge \mc{R}_{cd})}
    + 2\epsilon^{abcd}\delta \Gamma_{ab}\wedge D\mc{R}_{cd} \\
    &= \dd{\f(2\epsilon^{abcd}\delta \Gamma_{ab}\wedge \mc{R}_{cd})},
\end{split}
\end{equation}
where we used the property $D\epsilon^{abcd} = 0$ which follows from metric compatibility $\Gamma_{ab} = -\Gamma_{ba}$ and the Bianchi identity $D\t{\mc{R}}{^a_b} = 0$. Let us also define
\begin{equation}
    \Theta_{\mc{E}}(\theta; \delta\theta) \coloneqq 2\epsilon^{abcd}\delta \Gamma_{ab}\wedge \mc{R}_{cd},
\end{equation}
so that $\delta \mc{E} = \dd{\Theta_{\mc{E}}}$. The remaining part of the Cartan-Yang-Mills Lagrangian is
\begin{equation}
\begin{split}
    L_1(\theta) &\coloneqq L_{\mr{CYM}} - \r{1}{4}\mc{E} \\
    &= \r{1}{4}\epsilon^{abcd}\f(C_{ab}\wedge C_{cd} - \mc{R}_{ab}\wedge\mc{R}_{cd}) \\
    &= \epsilon^{abcd}\theta_a\wedge P_b \wedge C_{cd}
    - \epsilon^{abcd}\theta_a\wedge P_b \wedge \theta_c\wedge P_d \\
    &= -\epsilon^{abcd}P_a\wedge P_b \wedge \theta_c\wedge \theta_d \\
    &= -4 \t{P}{^{[a}_a}\t{P}{^{b]}_b}\mr{Vol}.
\end{split}
\end{equation}
That decomposition is not unique; however, this is the one that will work for us and provide a suitable decomposition of our symplectic current potential. 
  
Because of (\ref{euler-variation}), $L_1$ gives the same equations of motion as $L_{\mr{CYM}}$, which are equivalent to the vanishing of the Bach tensor -- see (\ref{deltaL}). The symplectic current potential  $\Theta_{\rm CYM}$ (\ref{ThetaCYM1}), on the other hand, splits into two parts:
\begin{equation}
    \Theta_{\mr{CYM}} = \r{1}{4}\Theta_{\mc{E}} + \Theta_1,
\end{equation}
where
\begin{equation}\label{theta1-derivation}
\begin{split}
    \Theta_1(\theta;\delta\theta) &\coloneqq \Theta_{\mr{CYM}}(\theta;\delta\theta) - \r{1}{4}\Theta_{\mc{E}}(\theta;\delta\theta) \\
    &= 2\delta\theta^a\wedge \hod DP_a + \r{1}{2}\epsilon^{abcd}\delta\Gamma_{ab}\wedge\f(C_{cd} - \mc{R}_{cd}) \\
    &= 2\delta\theta^a\wedge \hod DP_a + \epsilon^{abcd}\delta\Gamma_{ab}\wedge \theta_c \wedge P_d.
\end{split}
\end{equation}
From the decomposition (\ref{lcym-euler-decomposition}) and variations (\ref{deltaL}), (\ref{euler-variation}) follows
\begin{equation}
\delta L_1(\theta) = 2\delta \t{\theta}{^a}\wedge {B}_{ab}{\hod \theta}^b + \dd{\Theta_1}.
\end{equation}
Hence, $\Theta_1$ is a possible choice of the presymplectic potential current associated with the Lagrangian $L_1$.

Let us compare $\Theta_1$ with a presymplectic potential current $\Theta_{\mr{EH}}$ obtained from the Einstein-Hilbert Lagrangian $\hod \r{1}{16\pi G}\f(R - 2\Lambda)$, which can also be written as
\begin{equation}
    L_{\mr{EH}} = \r{1}{16\pi G}\f(\r{1}{2}\epsilon^{abcd}\theta_a\wedge\theta_b\wedge\mc{R}_{ab} - 2\Lambda \mr{Vol}).
\end{equation}
We have
\begin{equation}\label{eh-variation}
\begin{split}
    16\pi G \delta L_{\mr{EH}}(\theta) &= \r{1}{2}\epsilon^{abcd}\f(2\delta \theta_a \wedge \theta_b \wedge \mc{R}_{cd} + \theta_a\wedge \theta_b \wedge \delta\mc{R}_{cd}) - 2\Lambda \delta\theta_a \wedge \hod \theta^a \\
    &= \r{1}{2}\epsilon^{abcd}\f(2\delta \theta_a \wedge \theta_b \wedge \mc{R}_{cd} + \theta_a\wedge \theta_b \wedge D\delta\Gamma_{cd}) - 2\Lambda \delta\theta_a \wedge \hod \theta^a  \\
    &= \delta \theta_a\wedge \f(\epsilon^{abcd} \theta_b \wedge \mc{R}_{cd} - 2\Lambda \hod \theta^a) + \dd{\f(\r{1}{2}\epsilon^{abcd}\theta_a\wedge \theta_b \wedge \delta\Gamma_{cd})},
\end{split}
\end{equation}
where in the last equality we used the property $D\theta^a = 0$ which follows from $\t{\Gamma}{^a_b}$ being torsion-free. One can check that \cite{straumann_general_2004}
\begin{equation}
    \r{1}{2}\epsilon^{abcd}\theta_b\wedge\mc{R}_{cd} = \f(R \eta^{ab} - 2R^{ab})\hod \theta_b;
\end{equation}
therefore, the first term in (\ref{eh-variation}) yields the vacuum Einstein equations, while the second gives the presymplectic potential current \cite{depaoli_gaugeinvariant_2018}
\begin{equation}
    \Theta_{\mr{EH}}(\theta;\delta\theta) = \r{1}{32\pi G}\epsilon^{abcd}\theta_a\wedge\theta_b\wedge \delta\Gamma_{cd}.
\end{equation}
If $g = \eta_{ab}\theta^a\otimes\theta^b$ satisfies the vacuum Einstein equations, we have $P_a = P_{ab}\theta^b = -\r{\Lambda}{6}\theta_a$. Therefore, $DP^a = 0$ and $\Theta_1$ from (\ref{theta1-derivation}) reduces to
\begin{equation}
   \eval{\Theta_1(\theta;\delta\theta)}_{\mr{EH}} = -\r{\Lambda}{6}\epsilon^{abcd}\theta_a\wedge\theta_b\wedge \delta\Gamma_{cd},
\end{equation}
and hence, for those tetrads, $\Theta_1 = -\r{16\pi G \Lambda}{3}\Theta_{\mr{EH}}$.
Therefore, on solutions of the vacuum Einstein's equations,
\begin{equation}\label{theta-euler-splitting}
    \Theta_{\mr{CYM}} = \r{1}{4}\Theta_{\mc{E}} - \r{16\pi G \Lambda}{3}\Theta_{\mr{EH}}.
\end{equation}
In this way, on the Einstein spacetimes, $\Theta_{\mc{E}}$ can be thought of as a ``correction'' to the Einstein-Hilbert presymplectic potential which makes it conformally invariant.

\subsection{Noether currents of \texorpdfstring{$L_{\mr{CYM}}$}{the Cartan-Yang-Mills Lagrangian}}\label{ssec-noether-current}

Let $L(\Phi)$ be a Lagrangian form describing a general theory of fields $(\Phi_i)_{i\in I}$. Its variation with respect to the fields gives rise to the equations of motion and the presymplectic potential current
\begin{equation}\label{ncur-lag-var}
    \delta L(\Phi) = E(\Phi)_i\delta\phi^i + \dd{\Theta(\Phi;\delta\Phi)}.
\end{equation}
A variation $\delta_{\rm S}$ is a symmetry of the theory if the corresponding variation of the Lagrangian is an exact form
\begin{equation}\label{ncur-lag-symmetry}
    \delta_{\rm S}L(\Phi) = \dd{Z_{\rm S}(\Phi)}.
\end{equation}
For each such symmetry, we can define an associated Noether current \cite{iyer_properties_1994, compere_advanced_2019}
\begin{equation}\label{noether-current}
    J_{\rm S}(\Phi) = \Theta(\Phi;\delta_{\rm S}\Phi) - Z_{\rm S}(\Phi).
\end{equation}
If $\Phi_0$ satisfies the equations of motion, $E(\Phi_0) = 0$, from (\ref{ncur-lag-var}) and (\ref{ncur-lag-symmetry}) follows that $\dd{J_{\rm S}(\Phi_0)} = 0$ for any symmetry $\delta_{\rm S}$ of $L$. The Noether current is determined up to the addition of an exact three-form \cite{Wald1990} and depends on the particular choice of the presymplectic potential and $Z$.
If we consider a gauge transformation defined by a spacetime-dependent parameter $\lambda$, then by results of \cite{Wald1990}, as explained in \cite{iyer_properties_1994},
\begin{equation}
    J_\lambda(\theta) = \dd{Q_\lambda} 
\end{equation}
for some two-form $Q_\xi$ when the background satisfies the variational equations.
The form $Q_\lambda$, called the Noether charge, is not uniquely determined by $J_\xi$, since one can freely add to it any closed two-form.

Let us now turn to our Lagrangian $L_{\mr{CYM}}$. First, we will calculate the Noether current associated with the pseudorotations and conformal rescalings of the tetrad $\theta^a$. Under such transformation we have $\delta L = 0$; therefore, we only need to consider the presymplectic potential part of (\ref{noether-current}). As explained in Sec. \ref{sec-transformations}, the effects of both pseudorotations and conformal rescalings on the Cartan connection $\t{A}{^I_J}$ can be encoded in a subgroup of the group $SO(p+1,q+1)$ of matrices preserving the quadratic form (\ref{form-q}) according to the transformation law (\ref{homegah}). Hence, the variation of the connection can be expressed by the element of the algebra $\mf{so}(p+1,q+1)$ generating the particular transformation. Let us denote this generator by $\t{\gamma}{^I_J}$. Then
\begin{equation}
\begin{split}
    \delta_\gamma \t{A}{^I_J} &= \eval{\dv{t}}_{t=0}\f(\t{\exp(-\gamma t)}{^I_K}\t{A}{^K_L}\t{\exp(\gamma t)}{^L_J} + \t{\exp(-\gamma t)}{^I_K}\dd{\t{\exp(\gamma t)}{^K_J}}) \\
    &= \t{A}{^I_K}\t{\gamma}{^K_J} - \t{\gamma}{^I_K}\t{A}{^K_J} + \dd{\t{\gamma}{^I_J}} \\
    &= D_A \t{\gamma}{^I_J}.
\end{split}
\end{equation}
Thus
\begin{equation}\label{current-conformal-rotations}
\begin{split}
    J_\gamma(\theta) &= \Theta_{\mr{CYM}}(\theta;\delta_\gamma \theta) \\
    &= D_A \t{\gamma}{^I_J}\wedge\hod \t{F}{^J_I} \\
    &= \dd{\f(\t{\gamma}{^I_J}\wedge\hod \t{F}{^J_I})} + \t{\gamma}{^I_J}\wedge D_A\hod\t{F}{^J_I}.
\end{split}
\end{equation}
In the case when $\gamma$ generates a conformal rescaling $\theta^a \mapsto \exp(\alpha t)\theta^a$, which is when the matrix $\exp(\gamma t)$ is of the form (\ref{h-conformal}), one can check that this current vanishes identically (including off-shell):
\begin{equation}\label{current-conformal}
    J_\alpha(\theta) = 0.
\end{equation}
This follows from (\ref{fc}), (\ref{D*F}), and the fact that such matrices $\gamma$ have the form
\begin{equation}
    \gamma = \m{ *&0&0 \\ *&0&0 \\ *&*&* }.
\end{equation}
On the other hand, in the case when $\gamma$ generates a matrix representing a pseudorotation of the form (\ref{h-lorentz}), we obtain
\begin{equation}\label{current-rotations}
\begin{split}
    J_\omega (\theta) = \dd{\f(\t{\omega}{^a_b}\wedge\hod \t{C}{^b_a})},
\end{split}
\end{equation}
where $\omega \in \mf{so}(p,q)$ is the generator of the pseudorotation. Here again we use (\ref{fc}), (\ref{D*F}), as well as the form of the matrix $\gamma$:
\begin{equation}
    \gamma = \m{ 0&0&0 \\ 0&\t{\omega}{^a_b}&0 \\ 0&0&0 }.
\end{equation}
While not zero, this current is conserved for all tetrads, not only those satisfying the equations of motion $D_A \hod F = 0$.

Next, we will calculate the Noether current associated with a diffeomorphism generated by a vector field $\xi$. Since our theory is diffeomorphically invariant, the variation of $L_{\mr{CYM}}$ induced by a variation $\delta_\xi \theta = \lie_\xi \theta$ is simply the Lie derivative of $L_{\mr{CYM}}$:
\begin{equation}
    \delta_\xi L_{\mr{CYM}}(\theta) = \lie_\xi L_{\mr{CYM}}(\theta) = \xi \ipr L_{\mr{CYM}}(\theta)
    = \xi \ipr \f(\r{1}{2}\t{F}{^I_J}\wedge\hod\t{F}{^J_I}),
\end{equation}
where in the second equality we used the fact that the Lagrangian is a top form.
Using the expression for the presymplectic potential current $\Theta_{\mr{CYM}}$ in terms of the normal conformal Cartan connection and its curvature (\ref{cym-lagrangian}), (\ref{ThetaCYM1}) we obtain from (\ref{noether-current})
\begin{equation}\label{cnc-noether-current1}
    J_\xi(\theta) = \lie_\xi \t{A}{^I_J}\wedge \t{\hod F}{^J_I} - \xi \ipr \f(\r{1}{2}\t{F}{^I_J}\wedge \t{\hod F}{^J_I}).
\end{equation}
Note that this current is conformally invariant.
Next, we use the fact that the left and right duals of the Weyl tensor coincide \cite{batista_weyl_2013} [see also (\ref{weyl-dual})] to show that
\begin{equation}\label{xiff}
\begin{split}
    \xi \ipr \f(\r{1}{2}\t{F}{^I_J}\wedge \t{\hod F}{^J_I})
    &= \xi \ipr \f(\r{1}{2}\t{C}{^a_b}\wedge \t{\hod C}{^b_a}) \\
    &= \xi \ipr \f(\r{1}{4}\t{\epsilon}{^b_a_c^d}\t{C}{^a_b}\wedge \t{C}{^c_d}) \\
    &= \r{1}{2}\t{\epsilon}{^b_a_c^d}\f(\xi \ipr \t{C}{^a_b})\wedge \t{C}{^c_d} \\
    &= \f(\xi \ipr \t{F}{^I_J})\wedge \t{\hod F}{^J_I}
\end{split}
\end{equation}
and hence,
\begin{equation}
\begin{split}
    J_\xi(\theta) &= \f(\lie_\xi \t{A}{^I_J} - \xi \ipr \t{F}{^I_J}) \wedge \t{\hod F}{^J_I} \\
    &= \f(\dd{\f(\xi \ipr \t{A}{^I_J})} + \xi \ipr \dd{\t{A}{^I_J}} - \xi \ipr
    \f(\dd{\t{A}{^I_J}} + \t{A}{^I_K}\wedge\t{A}{^K_J}) ) \wedge \t{\hod F}{^J_I} \\
    &= \f(\dd{\f(\xi \ipr \t{A}{^I_J})} - \t{A}{^K_J}\f(\xi \ipr \t{A}{^I_K}) + \t{A}{^I_K}\f(\xi \ipr \t{A}{^K_J})) \wedge \t{\hod F}{^J_I} \\
    &= D_A \f(\xi \ipr \t{A}{^I_J}) \wedge \t{\hod F}{^J_I}.
\end{split}
\end{equation}
Integrating by parts we obtain the decomposition
\begin{equation}\label{noether-current-lie}
    J_\xi(\theta) = \dd{\f(\f(\xi \ipr \t{A}{^I_J}) \t{\hod F}{^J_I})} - \f(\xi \ipr \t{A}{^I_J}) \t{D_A \hod F}{^J_I} \eqqcolon \dd{Q_\xi} + \xi^a C_a,
\end{equation}
where $C_a = - \t{A}{^I_J_a} \t{D\hod F}{^J_I}$ are constraints that vanish whenever the equations of motion $D\hod F = 0$ hold and
\begin{equation}\label{noether-charge-lie}
    Q_\xi(\theta) = \f(\xi \ipr \t{A}{^I_J})\t{\hod F}{^J_I}
\end{equation}
is a possible choice for the Noether charge associated with $\xi$. Using it, we can also calculate the associated Iyer-Wald charge Hamiltonian charge \cite{iyer_properties_1994, compere_advanced_2019, kolanowski_hamiltonian_2021}
\begin{equation}
\begin{split}
    \cancel{\delta} H_\xi(\theta,\delta\theta) &= \delta Q_\xi(\theta) - \xi \ipr \Theta_{\mr{CYM}}(\theta;\delta \theta) \\
    &= \f(\xi \ipr \t{A}{^I_J})\delta\f(\hod \t{F}{^J_I})
    + \delta \t{A}{^I_J} \wedge \f(\xi \ipr \hod \t{F}{^J_I}).
\end{split}
\end{equation}

Let us examine the behavior of this current under pseudorotations and conformal rescalings of the tetrad. Let $\theta^a \mapsto \t{T}{^a_b}\theta^b$ be such a transformation, that is, $\t{T}{^a_b}\theta^b = f\theta^a$ for $f \in C^\infty(M)$ or $\t{T}{^a_b} \in SO(p,q)$. Then we have
\begin{equation}\label{jxi-transformation}
\begin{split}
    J_\xi(\t{T}{^a_b}\theta^b) &= \Theta_{\mr{CYM}}(\t{T}{^a_b}\theta^b; \lie_\xi\f(\t{T}{^a_b}\theta^b)) - \lie_\xi L_{\mr{CYM}}(\t{T}{^a_b}\theta^b) \\
    &= \Theta_{\mr{CYM}}(\t{T}{^a_b}\theta^b; \t{T}{^a_b}\lie_\xi\theta^b)
    + \Theta_{\mr{CYM}}(\t{T}{^a_b}\theta^b; \xi(\t{T}{^a_b}) \theta^b)- \lie_\xi L_{\mr{CYM}}(\t{T}{^a_b}\theta^b) \\
    &= \Theta_{\mr{CYM}}(\theta^b; \lie_\xi\theta^b)
    + \Theta_{\mr{CYM}}(\theta^b; \t{\f(T^{-1})}{^a_c}\xi(\t{T}{^c_b}) \theta^b)- \lie_\xi L_{\mr{CYM}}(\theta^b) \\
    &= J_\xi(\theta^a) + J_{T^{-1}\xi(T)}(\theta^a),
\end{split}
\end{equation}
where $J_{T^{-1}\xi(T)}$ is the current $J_\gamma$ from (\ref{current-conformal-rotations}), where $\gamma$ is the element of $\mf{so}(p+1,q+1)$ corresponding to the generator of pseudorotations/rescalings $T^{-1}\xi(T)$. This, together with (\ref{current-conformal}) and (\ref{current-rotations}) means that $J_\xi(\theta)$ is invariant under conformal rescaling of the tetrad, while under a pseudorotation it changes by the exact three-form (\ref{current-rotations}).

One could also try to obtain an invariant current by using a covariant Lie derivative \cite{depaoli_gaugeinvariant_2018}
\begin{equation}\label{cov-lie-tetrad}
    \lieg_\xi \theta^a \coloneqq \lie_\xi \theta^a + \f(\xi \ipr \t{\Gamma}{^a_b})\theta^b.
\end{equation}
The geometric meaning of this operation is that we lift the vector field $\xi$ to the frame bundle of $M$ horizontally with respect to $\t{\Gamma}{^a_b}$, consider the pullback of the differential forms $\theta^a$ to this frame and calculate the Lie derivative on the bundle instead of on the base space $M$. The variation of the tetrad connection associated with $\delta\theta^a = \lieg_\xi \theta^a$ is
\begin{equation}
    \lieg_\xi \t{\Gamma}{^a_b} =
    \lie_\xi \t{\Gamma}{^a_b} + \f(\xi \ipr \t{\Gamma}{^a_c})\t{\Gamma}{^c_b} - \f(\xi \ipr \t{\Gamma}{^c_b})\t{\Gamma}{^a_c} = \xi \ipr \f(\r{1}{2}\t{R}{^a_b_c_d}\theta^c\wedge \theta^d),
\end{equation}
which one can verify either by considering the tetrad connection as a one-form on the frame bundle or by directly checking that
\begin{equation}
    \dd{\f(\lieg_\xi \theta^a)} + \t{\Gamma}{^a_b}\wedge\f(\lieg_\xi \theta^b) +  \f(\lieg_\xi \t{\Gamma}{^a_b})\wedge\theta^b = 0,
    \quad \eta_{ac}\lieg_\xi \t{\Gamma}{^c_b} = -\eta_{bc}\lieg_\xi \t{\Gamma}{^c_a}.
\end{equation}
Consider now a pair of orthonormal tetrads $\theta^a$ and $\theta^{\prime a} = \t{T}{^a_b}\theta^b$. The corresponding tetrad connections are related by the transformation
\begin{equation}
    \t{\Gamma}{^\prime^a_b} = \t{T}{^a_c}\t{\Gamma}{^c_d}\t{\f(T^{-1})}{^d_b} - \t{\f(T^{-1})}{^c_b}\dd{\t{T}{^a_c}}.
\end{equation}
Using (\ref{cov-lie-tetrad}), we obtain
\begin{equation}
    \liegp_\xi \theta^{\prime a} = \t{T}{^a_b} \lieg_\xi \theta^b.
\end{equation}
Hence, the presymplectic potential current $\Theta_{\mr{CYM}}$ associated with the variation $\delta\theta^a = \lieg_\xi \theta^a$ is invariant with respect to arbitrary (pseudo)rotations and conformal rescalings. Moreover, since the modified Lie derivative $\lieg_\xi$ agrees with $\lie_\xi$ on objects without free tetrad indices, we have
\begin{equation}
    \lieg_\xi L_{\mr{CYM}}(\theta) = \lie_\xi L_{\mr{CYM}}(\theta) = 
    \dd{\f(\xi \ipr \r{1}{2}\t{F}{^I_J}\wedge\t{\hod F}{^J_I})}
    = \dd{\f(\f(\xi \ipr \t{F}{^I_J})\wedge\t{\hod F}{^J_I})}
\end{equation}
(see \ref{xiff}). Hence, the associated Noether current
\begin{equation}\label{current-diff-cov}
\begin{split}
    \ti{J}_\xi (\theta) &\coloneqq \Theta_{\mr{CYM}}(\theta;\lieg_\xi \theta) - \lieg_\xi L_{\mr{CYM}}(\theta) \\
    &= 2\lieg_\xi \theta^a \wedge \hod DP_a + \f(\lieg_\xi \t{\Gamma}{^a_b} - \xi \ipr \t{C}{^a_b})\wedge\t{\hod C}{^b_a}
\end{split}
\end{equation}
depends only on $g = \eta_{ab}\theta^a\theta^b$ and is conformally invariant. Since
\begin{equation}
    \lieg_\xi \t{\Gamma}{_a_b} - \xi \ipr \t{C}{_a_b} 
    = \xi \ipr \f(\mc{R}_{ab} - C_{ab})
    = \xi \ipr 2\theta_a\wedge P_b
    = 2\theta_{[a}P_{b]c}\xi^c - 2\xi_{[a}P_{b]}
\end{equation}
and
\begin{equation}
    \lieg_\xi \theta^a = \lie_\xi \theta^a + \f(\xi \ipr \t{\Gamma}{^a_b})\theta^b
    = \dd{\xi^a} + \xi \ipr \dd{\theta^a} + \f(\xi \ipr \t{\Gamma}{^a_b})\theta^b
    = D\xi^a,
\end{equation}
we can rewrite the current (\ref{current-diff-cov}) as
\begin{equation}
\begin{split}
    \ti{J}_\xi (\theta) &= 2D\xi^a \wedge \hod P_a - 2\xi^a P_b \wedge\t{\hod C}{^b_a} \\
    &= \dd{\f(2\xi^a \hod DP_a)} - 2\xi^a B_{ab} \hod \theta^b \\
    &\eqqcolon \dd{\ti{Q}_\xi} - \xi^a \ti{C}_a,
\end{split}
\end{equation}
where
\begin{equation}
    \ti{Q}_\xi \coloneqq 2\xi^a \hod DP_a
\end{equation}
is the Noether charge and $\ti{C}_a \coloneqq B_{ab}\hod \theta^b$ are constraints that vanish when the Bach tensor vanishes. Note that on the solutions of the vacuum Einstein equations $DP_a = 0$ and thus the current vanishes.

\section{\label{sec-boundary}Cartan-Yang-Mills presymplectic potential current on the conformal boundary}

\subsection{Asymptotically de Sitter spacetimes}\label{sec-ds}
In this section, we  restrict our considerations to metric tensors  of the signature $(-,+,+,+)$ on the manifold $M$.
Moreover, we assume that the corresponding spacetime is asymptotically de Sitter, and therefore the cosmological constant is positive throughout this section,       
\begin{equation}
 \Lambda\ >\ 0 .   
\end{equation}

A metric tensor $g$ that is asymptotically de Sitter  can be written (or defined) in the Fefferman-Graham gauge, that is  \cite{compere_lbms4_2019},
\begin{equation}\label{fefferman}
    g = \r{\ell^2}{\rho^2}\f(-\dd{\rho}^2 + \sum_{n=0}^\infty \rho^n g^{(n)}_{ij}(x^1,x^2,x^3)\dd{x^i}\dd{x^j}),
\end{equation}
where the asymptotic expansion amounts to expanding in $\rho>0$ around the boundary $\scri$ defined by the equation
\begin{equation}
\rho=0,
\end{equation}
contained in the conformal completion of $(M,g)$. 
The goal of this section is to calculate the symplectic current potential $\Theta_{\rm CYM}$ at $\scri$. By comparison, the symplectic current potential $\Theta_{\rm EH}$ is known to be ill-defined in that limit. However, it follows from  the conformal invariance that $\Theta_{\rm CYM}$ is well defined at $\scri$.   

It will be convenient to use a conformally rescaled metric tensor that is finite at $\rho=0$:
\begin{equation}\label{fefferman-ghat}
    \hat{g} \coloneqq \r{\rho^2}{\ell^2}g = -\dd{\rho}^2 + \sum_{n=0}^\infty \rho^n g^{(n)}_{ij}\dd{x^i}\dd{x^j}
\end{equation}
and its inverse 
\begin{equation}
    \hat{g}^{\alpha\beta}\partial_\alpha\otimes\partial_\beta = -\partial_\rho\otimes \partial_\rho + \sum_{n=0}^\infty \rho^n g^{ij}_{(n)}\pd_i \otimes\pd_j,
\end{equation}
where the matrix $g^{ij}_{(0)}$ is the inverse of $g^{(0)}_{ij}$, while
\begin{equation}\label{inverse}
    g^{ij}_{(n)} = - g_{(0)}^{ik}\sum_{m=1}^n g^{(m)}_{kl}g_{(n-m)}^{lj}.
\end{equation}
We will use the metric $\hat{g}$ and its inverse to raise and lower indices $\alpha,\beta,\gamma,\ldots$ on tensors with a hat and $g$ on the unhatted ones.

We will denote the pullback of the rescaled metric $\hat{g}_{\alpha\beta}$ to $\scri$ by $\mri{g}_{ij}$. If $\hat{g}_{\alpha\beta}$ has the Fefferman-Graham form (\ref{fefferman-ghat}), we have
\begin{equation}
 \mri{g}_{ij} = g^{(0)}_{ij}.   
\end{equation}
Einstein's equations (\ref{Ee}) imposed on the metric tensor (\ref{fefferman-ghat}) imply \cite{kolanowski_hamiltonian_2021, compere_lbms4_2019}
\begin{equation}
    g_{ij}^{(1)} = 0,
    \qquad g_{ij}^{(2)} = \mri{R}_{ij} - \r{1}{4}\mri{g}_{ij}\mri{R} =:  \mri{S}_{ij}, \qquad  g^{(0)}{}^{ij}g^{(3)}_{ij} = \mri{D}^i g^{(3)}_{ij} = 0,
\end{equation}
where $\mri{S}_{ij}$, $\mri{R}_{ij}$, $\mri{R}$ are, respectively, the Schouten tensor, Ricci tensor and Ricci scalar of $\mri{g}_{ij}$, and $\mri{D}_i$ is the metric-compatible, torsion-free connection defined by the metric tensor $\mri{g}_{ij}$. We will also use the notation 
\begin{equation}
\mri{T}_{ij} := g^{(3)}_{ij}. 
\end{equation}
Equation (\ref{inverse}) implies
\begin{equation}
    g^{ij}_{(1)} = 0,
    \qquad g^{ij}_{(2)} = -\mri{S}^{ij},
    \qquad g^{ij}_{(3)} = -\mri{T}^{ij}.
\end{equation}
(Indices $i,j,\ldots$ on $\mri{S}_{ij}$ and $\mri{T}_{ij}$ are raised with $\mri{g}^{ij}$ and lowered with $\mri{g}_{ij}$).

Let us consider a coframe $\theta^a$ such that
\begin{equation}
    g = \eta_{ab}\theta^a\theta^b,
\end{equation}
where $\eta_{ab} = \diag\f(-,+,+,+)$. We will denote its dual frame by $e_a$.
From (\ref{fefferman}) it follows that for any such tetrad $\theta^a = \r{\ell}{\rho}\hat{\theta}^a$ and $e_a = \r{\rho}{\ell}\hat{e}_a$, where $\hat{\theta}^a$ and $\hat{e}_a$ are the orthonormal coframe and the dual frame associated with the rescaled metric $\hat{g}$ from (\ref{fefferman-ghat}). Let us calculate the pullback of $\Theta_{\mr{CYM}}$ to $\scri$ for this particular choice of $\theta^a$.
Since we are considering an Einstein spacetime (\ref{Ee}), we can use the formula (\ref{ThetaCYM3})
\begin{equation}\label{theta-ym-simple}
    \Theta_{\mr{CYM}}(\theta^a,\delta\theta^a) = \delta\t{\Gamma}{^b_c}\wedge\t{*C}{^c_b}.
\end{equation}
The tetrad connection of $\theta^a$ is \cite{krasnov_formulations_2020}
\begin{equation}\label{gamma-ab}
    \t{\Gamma}{^a_b} = \eta^{ac}e_{c}^\alpha e_{b}^\beta \f(c_{\alpha\beta\gamma} + c_{\beta\gamma\alpha} - c_{\gamma\alpha\beta})\dd{x^\gamma},
\end{equation}
where
\begin{equation}\label{bound-cab}
    c_{\alpha\beta\gamma} = \eta_{ab}\theta^a_\alpha \pd_{[\beta}\theta^b_{\gamma]}.
\end{equation}
Let us expand the one-forms $\t{\Gamma}{^a_b}$ in terms of $\rho$. First of all, since $\theta^a = \r{\ell}{\rho}\hat{\theta}^a$ and $e_a = \r{\rho}{\ell} \hat{e}_a$, we have
\begin{equation}
    c_{\alpha\beta\gamma} = -\r{\ell^2}{\rho^3} \eta_{ab}\hat{\theta}^a_\alpha \t{\delta}{^\rho_{[\beta}}\hat{\theta}^b_{\gamma]} + \order*{\rho^{-2}}
    = -\r{\ell^2}{\rho^3}\t{\delta}{^\rho_{[\beta}}\hat{g}_{\gamma]\alpha} + \order*{\rho^{-2}}
\end{equation}
and from (\ref{gamma-ab}) follows
\begin{equation}\label{gamma-tetrad-rho}
\begin{split}
    \t{\Gamma}{^a_b} &= -\rho^{-1}\eta^{ac}\hat{e}_{c}^\alpha \hat{e}_{b}^\beta \f(\t{\delta}{^\rho_{[\beta}}\hat{g}_{\gamma]\alpha} + \t{\delta}{^\rho_{[\gamma}}\hat{g}_{\alpha]\beta} - \t{\delta}{^\rho_{[\alpha}}\hat{g}_{\beta]\gamma})\dd{x^\gamma} + \order*{1} \\
    &= 2\rho^{-1}\eta^{ac}\hat{e}_{c}^\alpha \hat{e}_{b}^\beta \t{\delta}{^\rho_{[\alpha}}\hat{g}_{\beta]\gamma}\dd{x^\gamma} + \order*{1} \\
    &= 2\rho^{-1}\eta^{ac}\hat{e}_{[c}^\alpha \hat{e}_{b]}^\beta \t{\delta}{^\rho_{\alpha}}\hat{g}_{\beta\gamma}\dd{x^\gamma} + \order*{1} \\
    &= 2\rho^{-1}\eta^{ac}\hat{e}_{[c}^\rho \hat{e}_{b]}^\beta \hat{g}_{\beta\gamma}\dd{x^\gamma} + \order*{1}.
\end{split}
\end{equation}

Moreover, the Weyl tensor is conformally invariant,
\begin{equation}\label{weyl-transformation}
    \t{C}{^\alpha_\beta_\gamma_\delta} = \t{\hat{C}}{^\alpha_\beta_\gamma_\delta}
\end{equation}
and the relation between the volume forms of $\hat{g}_{\alpha\beta}$ and $g_{\alpha\beta}$ is as follows:
\begin{equation}
    \hat{\epsilon}_{\alpha\beta\gamma\delta} = \sqrt{\abs{\det\hat{g}}}\varepsilon_{\alpha\beta\gamma\delta}
    = \r{\rho^4}{\ell^4}\sqrt{\abs{\det g}}\varepsilon_{\alpha\beta\gamma\delta}
    = \r{\rho^4}{\ell^4}\epsilon_{\alpha\beta\gamma\delta}.
\end{equation}
Therefore,
\begin{equation}
    \t{\hat{\epsilon}}{^\alpha^\beta_\gamma_\delta} = \t{\epsilon}{^\alpha^\beta_\gamma_\delta}
\end{equation}
(since we use $\hat{g}^{\alpha\beta}$ to raise indices on the hatted quantities and $g^{\alpha \beta}$ -- on the unhatted). Consequently
\begin{equation}\label{weyl-star-transformation}
    \t{\hod C}{^\alpha_\beta} = \r{1}{4}\t{C}{^\alpha_\beta_\gamma_\delta}\t{\epsilon}{^\gamma^\delta_\epsilon_\zeta}\dd{x^\epsilon}\wedge\dd{x^\zeta}
    = \r{1}{4}\t{\hat{C}}{^\alpha_\beta_\gamma_\delta}\t{\hat{\epsilon}}{^\gamma^\delta_\epsilon_\zeta}\dd{x^\epsilon}\wedge\dd{x^\zeta}
    = \hat{\hod}\t{\hat{C}}{^\alpha_\beta}.
\end{equation}
Since
\begin{equation}
    \hat{g} = -\dd{\rho}^2 + \mri{g}_{ij}\dd{x^i}\dd{x^j} + \order*{\rho^2},
\end{equation}
we have
\begin{equation}
    \det \hat{g} = -\det \mri{g} + \order*{\rho^2},
\end{equation}
and thus
\begin{equation}
    \sqrt{\abs{\det\hat{g}}} = \sqrt{\det \mri{g}} + \order*{\rho^2}.
\end{equation}
Therefore
\begin{equation}\label{volume-forms-relation}
    \hat{\epsilon}_{\rho ijk} = \sqrt{\abs{\det\hat{g}}}\varepsilon_{\rho ijk}
    = \sqrt{\det\mri{g}}\varepsilon_{ijk} + \order*{\rho^2}
    = \mri{\epsilon}_{ijk} + \order*{\rho^2},
\end{equation}
where $\mri{\epsilon}_{ijk} \coloneqq \sqrt{\mri{g}}\varepsilon_{ijk}$ is a three-form defined on a neighborhood of $\scri$, whose pullback to $\scri$ is the volume form on $\scri$ induced by the conformal metric $\hat{g}$. As a consequence of (\ref{volume-forms-relation}), we have
\begin{equation}\label{weyl-star-fef}
    \r{1}{2}\t{\hat{C}}{^\beta_\alpha^\gamma^\delta}\hat{\epsilon}_{\gamma\delta jk}
    = \t{\hat{C}}{^\beta_\alpha^\rho^l}\mri{\epsilon}_{ljk} + \order*{\rho^2}
    = -\t{\hat{C}}{^\beta_\alpha_\rho^l}\mri{\epsilon}_{ljk} + \order*{\rho^2}.
\end{equation}

The pullback of (\ref{theta-ym-simple}) to $\scri$ is
\begin{equation}
    \ti{\Theta}_{\mr{CYM}} = \lim_{\rho\ar 0} \delta\t{\Gamma}{^a_b_i} \hat{e}_a^\alpha \hat{\theta}^b_\beta \dd{x^i} \wedge \f(\r{1}{2} \t{\hat{C}}{^\beta_\alpha^\gamma^\delta} \hat{\epsilon}_{\gamma\delta jk} \dd{x^j} \wedge \dd{x^k}).
\end{equation}
Using the formula for the variation of the tetrad connection (\ref{gamma-tetrad-rho}), we get
\begin{equation}
\begin{split}
    \ti{\Theta}_{\mr{CYM}} &= \lim_{\rho\ar 0} \delta\f(2\rho^{-1}\eta^{ac}\hat{e}_{[c}^\rho \hat{e}_{b]}^\zeta \hat{g}_{\zeta i}) \hat{e}_a^\alpha \hat{\theta}^b_\beta \dd{x^i} \wedge \f(\r{1}{2}\t{\hat{C}}{^\beta_\alpha^\gamma^\delta} \hat{\epsilon}_{\gamma\delta jk} \dd{x^j} \wedge \dd{x^k}).
\end{split}
\end{equation}
Next, we use (\ref{weyl-star-fef}) to obtain
\begin{equation}\label{theta-scri1}
\begin{split}
    \ti{\Theta}_{\mr{CYM}} &= -\lim_{\rho\ar 0} \delta\f(2\rho^{-1}\eta^{ac}\hat{e}_{[c}^\rho \hat{e}_{b]}^\zeta \hat{g}_{\zeta i}) \hat{e}_a^\alpha \hat{\theta}^b_\beta \dd{x^i} \wedge \f(\t{\hat{C}}{^\beta_\alpha_\rho^l} \mri{\epsilon}_{ljk} \dd{x^j} \wedge \dd{x^k}) \\
    &= -\lim_{\rho\ar 0} \delta\f(2\rho^{-1}\eta^{ac}\hat{e}_{[c}^\rho \hat{e}_{b]}^\zeta \hat{g}_{\zeta i}) \hat{e}_a^\alpha \hat{\theta}^b_\beta \t{\hat{C}}{^\beta_\alpha_\rho^i} \mri{\mr{Vol}} \\
    &= -\lim_{\rho\ar 0} \delta\f(2\rho^{-1}\hat{e}_{[a}^\rho \hat{e}_{b]}^\zeta \hat{g}_{\zeta i}) \hat{\theta}^a_\alpha \hat{\theta}^b_\beta \t{\hat{C}}{^\beta^\alpha_\rho^i} \mri{\mr{Vol}} \\
    &= \lim_{\rho\ar 0} \delta\f(2\rho^{-1}\hat{e}_{a}^\rho \hat{e}_{b}^\zeta \hat{g}_{\zeta i}) \hat{\theta}^a_\alpha \hat{\theta}^b_\beta \t{\hat{C}}{^\beta^\alpha^\rho^i} \mri{\mr{Vol}} \\
    &= \lim_{\rho\ar 0} 2\rho^{-1}\f(\delta\hat{e}_{a}^\rho \hat{e}_{b}^\zeta \hat{g}_{\zeta i} + \hat{e}_a^\rho \delta\hat{\theta}^c_i\eta_{cb}) \hat{\theta}^a_\alpha \hat{\theta}^b_\beta \t{\hat{C}}{^\beta^\alpha^\rho^i} \mri{\mr{Vol}} \\
    &= \lim_{\rho\ar 0} 2\rho^{-1}\f(\delta\hat{e}_{a}^\rho \hat{g}_{\beta i}\hat{\theta}^a_\alpha \t{\hat{C}}{^\beta^\alpha^\rho^i} +  \delta\hat{\theta}^c_i\eta_{cb} \hat{\theta}^b_\beta \t{\hat{C}}{^\beta^\rho^\rho^i}) \mri{\mr{Vol}},
\end{split}
\end{equation}
where $\mri{\mr{Vol}} \coloneqq \r{1}{3!}\mri{\epsilon}_{ijk}\dd{x^i}\wedge\dd{x^j}\wedge\dd{x^k}$. Using the fact that the Weyl tensor is traceless in every pair of indices, we get
\begin{equation}
    \hat{g}_{\beta i}\t{\hat{C}}{^\beta^\alpha^\rho^i} = \hat{g}_{\beta \gamma}\t{\hat{C}}{^\beta^\alpha^\rho^\gamma} - \hat{g}_{\beta\rho}\t{\hat{C}}{^\beta^\alpha^\rho^\rho} = -\hat{g}_{\beta\rho}\t{\hat{C}}{^\beta^\alpha^\rho^\rho} = 0.
\end{equation}
Thus the first term in the last line of (\ref{theta-scri1}) vanishes and we are left with
\begin{equation}
\begin{split}
    \tilde{\Theta}_{\mr{CYM}}(\theta;\delta\theta) &= \lim_{\rho\ar 0} 2\rho^{-1}  \delta\hat{\theta}^c_i\eta_{cb} \hat{\theta}^b_\beta \t{\hat{C}}{^\beta^\rho^\rho^i} \mri{\mr{Vol}} \\
    &= \lim_{\rho\ar 0} 2\rho^{-1} \delta\hat{\theta}^c_i\eta_{cb} \hat{\theta}^b_j \t{\hat{C}}{^j^\rho^\rho^i} \mri{\mr{Vol}} \\
    &= \lim_{\rho\ar 0}\rho^{-1}\delta\hat{g}_{ij}\t{\hat{C}}{^i^\rho^\rho^j}\mri{\mr{Vol}} \\
    &= \r{3}{2}\delta\mri{g}_{ij}\mri{T}^{ij}\mri{\mr{Vol}},
\end{split}
\end{equation}
where in the last step we used Eq. (\ref{fef-weyl-rirj}) from Appendix \ref{sec-fefferman}. Notice that although in general, the presymplectic potential current depends on the tetrad used to define the normal conformal Cartan connection and its variation, it turns out that at the conformal boundary of de Sitter spacetime, this dependence only manifests through the metric, $g = \eta_{ab}\theta^a\theta^b$, and the variation of the metric induced on $\scri$ by the conformally rescaled metric $\hat{g}$.

Moreover, note that here $\delta\mri{g}_{ij}$ means simply the variation of the intrinsic metric on $\scri$ induced by a general variation $\delta g = \r{\ell^2}{\rho^2}\delta\hat{g}$. It does not mean that $\delta g$ has to preserve the Fefferman-Graham gauge (in which case we could write $\delta \mri{g}_{ij} = \delta g^{(0)}_{ij}$).

The standard definition of the holographic energy-momentum tensor on the boundary is \cite{compere_lbms4_2020}
\begin{equation}\label{tab}
    T_{ij} = \r{3\ell}{16\pi G}\mri{T}_{ij},
\end{equation}
which implies
\begin{equation}
    \tilde{\Theta}_{\mr{CYM}} = \r{8\pi G}{\ell}\delta\mri{g}_{ij}T^{ij}\mri{\mr{Vol}}.
\end{equation}
On the other hand, the variation of the holographically renormalized Einstein-Hilbert action,
\begin{equation}
    S_{\mr{GR}} = \r{1}{16\pi G}\int_M \f(R - 2\Lambda)\mr{Vol} + \r{1}{16\pi G}\int_\scri \f(2K + \r{4}{\ell} - \mri{R})\mri{\mr{Vol}},
\end{equation}
yields the following presymplectic potential current on $\scri$ \cite{compere_lbms4_2020}:
\begin{equation}
    \tilde{\Theta}_{\mr{GR}} = -\r{\ell}{2}\delta\mri{g}_{ij}T^{ij}\mri{\mr{Vol}}.
\end{equation}
Hence,
\begin{equation}\label{theta-cym-gr-ds}
\ti{\Theta}_{\mr{CYM}} = -\r{16\pi G\Lambda}{3} \ti{\Theta}_{\mr{GR}}.
\end{equation}

\subsection{Asymptotically flat spacetimes}\label{sec-flat}

We describe an asymptotically flat spacetime using the Bondi-Sachs gauge (in coordinates $\Omega$, $u$, $x^A$, where $\r{1}{\Omega}$ is the luminosity distance) \cite{sachs_gravitational_1962, madler_bondisachs_2016, strominger_lectures_2018}:
\begin{equation}\label{bondi-metric}
\begin{gathered}
    g_{\Omega\Omega} = g_{\Omega A} = 0, \quad
    g_{uu} = -1 + 2M\Omega + \order*{\Omega^2}, \quad
    g_{AB} = \Omega^{-2}\gamma_{AB} + \Omega^{-1}C_{AB} + \order*{1}, \\
    g_{u\Omega} = \Omega^{-2} - \r{1}{16}C^{AB}C_{AB} + \order*{\Omega}, \quad
    g_{uA} = \r{1}{2}D^B C_{BA} + \order*{\Omega},
\end{gathered}
\end{equation}
\begin{equation}\label{bondi-metric-inverse}
\begin{gathered}
    g^{uu} = g^{uA} = 0, \quad
    g^{\Omega\Omega} = \Omega^4 - 2M\Omega^5 + \order*{\Omega^6}, \quad
    g^{AB} = \Omega^{2}\gamma^{AB} - \Omega^3 C^{AB} + \order*{\Omega^4}, \\
    g^{u\Omega} = \Omega^{2} + \r{1}{16}C^{AB}C_{AB}\Omega^4 + \order*{\Omega^5}, \quad
    g^{\Omega A} = -\r{1}{2}D_B C^{BA}\Omega^4 + \order*{\Omega^5}.
\end{gathered}
\end{equation}
Here $M$ is the Trautman-Bondi mass aspect, $N_{AB} \coloneqq \pd_u C_{AB}$ is the Bondi news tensor, $\gamma_{AB}$ is a unit sphere metric, $D$ is its covariant derivative, and indices $A,B,\ldots$ on those objects are raised with $\gamma^{AB}$. Similar to the asymptotically de Sitter case, let us define $\hat{g} \coloneqq \Omega^2 g$. Let us also use indices $i,j,\ldots$ to go over the three coordinates $u,x^A$ and define $\mri{g}$ to be the pullback of $\hat{g}$ to $\scri$ (i.e., the surface $\Omega = 0$).

From (\ref{bondi-metric}) and $\hat{g} \coloneqq \Omega^2 g$, we obtain
\begin{equation}\label{bondi-ghat}
    \hat{g} = 2\dd{u}\dd{\Omega} + \gamma_{AB}\dd{x^A}\dd{x^B} + \order*{\Omega},
\end{equation}
so the pullback of $\hat{g}$ to $\scri$, which we denote by $\mri{g}$, is
\begin{equation}
    \mri{g} = \gamma_{AB}\dd{x^A}\dd{x^b}.
\end{equation}
Therefore
\begin{equation}
    \det \hat{g} = -\det \mri{g} + \order*{\Omega},
\end{equation}
\begin{equation}
    \sqrt{\abs{\det\hat{g}}} = \sqrt{\det\gamma} + \order*{\Omega},
\end{equation}
\begin{equation}
    \hat{\epsilon}_{\Omega ijk} = \sqrt{\abs{\det\hat{g}}}\varepsilon_{\Omega ijk}
    = \sqrt{\det\gamma}\varepsilon_{ijk} + \order*{\Omega}
    = \mri{\epsilon}_{ijk} + \order*{\Omega},
\end{equation}
where, as in the previous section,
\begin{equation}
\mri{\mr{Vol}} \coloneqq \r{1}{3!}\mri{\epsilon}_{ijk}\dd{x^i}\wedge\dd{x^j}\wedge\dd{x^k}
\coloneqq \r{1}{3!}\sqrt{\det\gamma}\varepsilon_{ijk}\dd{x^i}\wedge\dd{x^j}\wedge\dd{x^k}
\end{equation}
is a three-form defined in the neighborhood of $\scri$, whose pullback to $\scri$ is the volume form induced on $\scri$ by $\hat{g}$.

Since our spacetime satisfies $R_{ab} = 0$, we can calculate $\Theta_{\mr{CYM}}$ using
\begin{equation}
    \Theta_{\mr{CYM}} = \delta\t{\Gamma}{^a_b}\wedge\t{\hod C}{^b_a}.
\end{equation}
The peeling theorem implies that $\t{C}{^\alpha_\beta_\gamma_\delta} = \t{\hat{C}}{^\alpha_\beta_\gamma_\delta} = \order*{\Omega}$. Also,
\begin{equation}\label{gamma-tetrad-omega}
\begin{split}
    \t{\Gamma}{^a_b} &= -\Omega^{-1}\eta^{ac}\hat{e}_{c}^\alpha \hat{e}_{b}^\beta \f(\t{\delta}{^\Omega_{[\beta}}\hat{g}_{\gamma]\alpha} + \t{\delta}{^\Omega_{[\gamma}}\hat{g}_{\alpha]\beta} - \t{\delta}{^\Omega_{[\alpha}}\hat{g}_{\beta]\gamma})\dd{x^\gamma} + \order*{1} \\
    &= 2\Omega^{-1}\eta^{ac}\hat{e}_{[c}^\Omega \hat{e}_{b]}^\beta \hat{g}_{\beta\gamma}\dd{x^\gamma} + \order*{1}.
\end{split}
\end{equation}
Hence
\begin{equation}
\begin{split}
    \ti{\Theta}_{\mr{CYM}} &= \lim_{\Omega\ar 0} \delta\f(2\Omega^{-1} \hat{e}_{a}^\Omega \hat{e}_{b}^\zeta \hat{g}_{\zeta i} \dd{x^i})\hat{\theta}^a_\alpha \hat{\theta}^b_\beta \wedge \f(\r{1}{2}\t{\hat{C}}{^\beta^\alpha^\gamma^\delta}\hat{\epsilon}_{\gamma\delta jk}\dd{x^j}\wedge\dd{x^k}) \\
    &= \lim_{\Omega\ar 0} 2\Omega^{-1} \f(\hat{\theta}^a_\alpha \delta\hat{e}_{a}^\Omega \hat{g}_{\beta i} + \t{\delta}{^\Omega_\alpha} \eta_{cb}\delta\hat{\theta}^c_i \hat{\theta}^b_\beta) \dd{x^i} \wedge \f(\t{\hat{C}}{^\beta^\alpha^\Omega^l}\mri{\epsilon}_{ljk}\dd{x^j}\wedge\dd{x^k}) \\
    &= \lim_{\Omega\ar 0} 2\Omega^{-1} \f(\hat{\theta}^a_\alpha \delta\hat{e}_{a}^\Omega \hat{g}_{\beta i} + \t{\delta}{^\Omega_\alpha} \eta_{cb}\delta\hat{\theta}^c_i \hat{\theta}^b_\beta) \t{\hat{C}}{^\beta^\alpha^\Omega^i} \mri{\mr{Vol}} \\
    &= \lim_{\Omega\ar 0} \Omega^{-1} \delta\hat{g}_{ji} \t{\hat{C}}{^j^\Omega^\Omega^i} \mri{\mr{Vol}} \\
    &= \lim_{\Omega\ar 0} \Omega^{-1} \delta\mri{g}_{ji} \t{\hat{C}}{^j_u_u^i} \mri{\mr{Vol}}.
\end{split}
\end{equation}
Inserting the components of the Weyl tensor calculated in Appendix \ref{sec-bs}, we obtain
\begin{equation}
    \ti{\Theta}_{\mr{CYM}} = \f[\delta\mri{g}_{uu}\f(\r{1}{4}C^{AB}N_{AB} + 2M) + \r{1}{2}\delta\mri{g}_{AB}\pd_u N^{AB}]\mri{\mr{Vol}}.
\end{equation}
In this case there is no contribution from the Einstein-Hilbert potential, since that part of the decomposition (\ref{theta-euler-splitting}) has a proportionality constant of $\Lambda$, and thus vanishes for asymptotically flat spacetimes.

In the special case when $\delta \mri{g} = \lie_\xi \mri{g}$ for some BMS vector $\xi$, we have
\begin{equation}
    \lie_\xi \mri{g} = 2\alpha \mri{g},
\end{equation}
so
\begin{equation}\label{theta-bms}
    \ti{\Theta}_{\mr{CYM}}(\lie_\xi \mri{g}) =  \alpha\gamma_{AB}\pd_u N^{AB}\mri{\mr{Vol}} = 0,
\end{equation}
since the Bondi news tensor is traceless.

\section{Summary}

Using the normal conformal Cartan connection as a gauge potential of a Yang-Mills theory allowed us to obtain a conformally invariant presymplectic potential current (\ref{ThetaCYM2}). Because of the fact that the Yang-Mills current of our theory encodes the Bach tensor of the underlying spacetime (\ref{D*F}), the theory's equations of motion are satisfied by any tetrad that generates a metric conformally equivalent to a solution to the vacuum Einstein field equations. This made it viable to use our potential current to derive Noether currents and charges (\ref{noether-current}), (\ref{current-conformal-rotations}), (\ref{noether-current-lie}) conserved on such tetrads. In particular, we showed that the current associated with diffeomorphisms is conformally invariant. Additionally, we described the way in which, on the solutions to the vacuum Einstein's equations, the potential decomposes into a part proportional to the Einstein-Hilbert presymplectic potential and the topological (Euler) term (\ref{theta-euler-splitting}). As an example of an application of this formalism, we calculated the presymplectic potential current induced by our theory on the conformal boundary of asymptotically de Sitter spacetime and found out that the result is proportional to the potential one obtains using a holographically renormalized gravity action (\ref{theta-cym-gr-ds}). This provides a mathematically elegant way of obtaining a symplectic structure on that boundary without resorting to renormalization techniques. On the other hand, the potential obtained at the null infinity of an asymptotically flat spacetime does not seem to have a clear physical relevance and in fact vanishes for variations induced by the BMS diffeomorphism group (\ref{theta-bms}), which is to be expected since in the case of $\Lambda = 0$ only the topological part of the decomposition (\ref{theta-euler-splitting}) remains.

\begin{acknowledgments}

The research J.L. leading to these results has received funding from the Norwegian Financial Mechanism 2014-2021 UMO-2020/37/K/ST1/02788.

\end{acknowledgments}

\appendix

\section{\label{sec-fefferman}ASYMPTOTICALLY DE SITTER SPACETIMES IN THE FEFFERMAN-GRAHAM GAUGE}

We consider the metric
\begin{equation}\label{fef-metric}
    g = \r{\ell^2}{\rho^2}\f(-\dd{\rho}^2 + \sum_{n=0}^\infty\rho^n g^{(n)}_{ij}\dd{x^i}\dd{x^j})
\end{equation}
with an inverse
\begin{equation}
    g^{\alpha \beta}\pd_\alpha\otimes\pd_\beta = \r{\rho^2}{\ell^2}\f(-\pd_\rho\otimes\pd_\rho + \sum_{n=0}^\infty \rho^n g_{(n)}^{ij} \pd_i\otimes\pd_j),
\end{equation}
where the matrix $g_{(0)}^{ij}$ is the inverse of $g^{(0)}_{ij}$ and for $n \ge 1$ we have
\begin{equation}
    g_{(n)}^{ij} = -g_{(0)}^{ik}\sum_{m=1}^n g^{(m)}_{kl}g_{(n-m)}^{lj}.
\end{equation}

Imposing vacuum Einstein's equations $R_{\alpha\beta} = \Lambda g_{\alpha\beta}$ on (\ref{fef-metric}) with $\Lambda = \r{3}{\ell^2}$ leads to the following constraints \cite{kolanowski_hamiltonian_2021, compere_lbms4_2019}:
\begin{equation}
    g^{(1)}_{ij} = 0, \qquad g^{(2)}_{ij} = \mri{R}_{ij} - \r{1}{4}\mri{R}g^{(0)}_{ij}, \qquad  g^{(0)}{}^{ij}g^{(3)}_{ij} = \mri{D}^i g^{(3)}_{ij} = 0,
\end{equation}
where $\mri{R}_{ij}$, $\mri{R}$ and $\mri{D}_i$ are, respectively, the Ricci tensor, Ricci scalar, and the metric-compatible, torsion-free covariant derivative of $g^{(0)}_{ij}$.

To calculate the potential $\Theta_{\mr{CYM}}$, we need the Weyl tensor of the metric $g_{\alpha\beta}$. Since it is conformally invariant, we will calculate the Weyl tensor of a simpler metric,
\begin{equation}\label{fef-ghat}
    \hat{g} = \r{\rho^2}{\ell^2}g = -\dd{\rho}^2 + \sum_{n=0}^\infty\rho^n g^{(n)}_{ij}\dd{x^i}\dd{x^j}.
\end{equation}
First, let us introduce the following notation to simplify expressions:
\begin{equation}
    \mri{g}_{ij} \coloneqq g^{(0)}_{ij},\quad
    \mri{S}_{ij} \coloneqq g^{(2)}_{ij} = \mri{R}_{ij} - \r{1}{4}\mri{R}\mri{g}_{ij},\quad
    \mri{T}_{ij} \coloneqq g^{(3)}_{ij}.
\end{equation}

From (\ref{fef-ghat}) we derive the following Christoffel symbols  (below $\mri{D}_i$ is the covariant derivative of $\mri{g}_{ij}$ and $\t{\mri{\Gamma}}{^i_j_k}$ are its Christoffel symbols):
\begin{equation}
    \t{\hat{\Gamma}}{^\rho_\rho_\rho} =
    \t{\hat{\Gamma}}{^\rho_\rho_i} =
    \t{\hat{\Gamma}}{^i_\rho_\rho} = 0,
\end{equation}
\begin{equation}
    \t{\hat{\Gamma}}{^\rho_i_j}
    = \rho\mri{S}_{ij} + \r{3}{2}\rho^2 \mri{T}_{ij} + \order*{\rho^3},
\end{equation}
\begin{equation}
    \t{\hat{\Gamma}}{^i_j_\rho}
    = \rho \t{\mri{S}}{^i_j} + \r{3}{2}\rho^2\t{\mri{T}}{^i_j} + \order*{\rho^3},
\end{equation}
\begin{equation}
\begin{split}
    \t{\hat{\Gamma}}{^i_j_k}
    &= \t{\mri{\Gamma}}{^i_j_k} + \rho^2\f(\mri{S}^{il}\t{\mri{\Gamma}}{_{ljk}} + \r{1}{2}\mri{g}^{il}\f(2\pd_{(j}\mri{S}_{k)l} - \pd_l \mri{S}_{jk})) + \order*{\rho^3} \\
    &= \t{\mri{\Gamma}}{^i_j_k} + \rho^2\f(\t{\mri{D}}{_{(j}}\t{\mri{S}}{_{k)}^i} - \r{1}{2}\mri{D}^i\mri{S}_{jk}) + \order*{\rho^3}.
\end{split}
\end{equation}

Next, we calculate the Riemann tensor:
\begin{equation}
    \t{\hat{R}}{^\rho_i_\rho_j} = \pd_\rho \t{\hat{\Gamma}}{^\rho_i_j} - \t{\hat{\Gamma}}{^\rho_k_j}\t{\hat{\Gamma}}{^k_\rho_i}
    = \mri{S}_{ij} + 3\rho \mri{T}_{ij} + \order*{\rho^2},
\end{equation}
\begin{equation}
\begin{split}
    \t{\hat{R}}{^\rho_i_j_k} &= 2\t{\pd}{_{[j}}\t{\hat{\Gamma}}{^\rho_{k]}_i}
    + 2\t{\hat{\Gamma}}{^\rho_l_{[j}}\t{\hat{\Gamma}}{^l_{k]}_i} \\
    &= \rho\f(\pd_j \t{\mri{S}}{_k_i} - \pd_k \t{\mri{S}}{_j_i}
    + \t{\mri{S}}{_l_j}\t{\mri{\Gamma}}{^l_k_i} - \t{\mri{S}}{_l_k}\t{\mri{\Gamma}}{^l_j_i}) + \order*{\rho^2} \\
    &= \rho\f(\mri{D}_j \t{\mri{S}}{_k_i} - \mri{D}_k \t{\mri{S}}{_j_i}) + \order*{\rho^2} \\
    &= 2\rho \mri{D}_{[j}\mri{S}_{k]i} + \order*{\rho^2},
\end{split}
\end{equation}
\begin{equation}
    \t{\hat{R}}{^i_j_k_l} = 2\pd_{[k} \t{\hat{\Gamma}}{^i_{l]}_j}
    + 2\t{\hat{\Gamma}}{^i_m_{[k}}\t{\hat{\Gamma}}{^m_{l]}_j}
    = \t{\mri{R}}{^i_j_k_l} + \order*{\rho^2}.
\end{equation}
Lowering the first index, we get the only (up to symmetries) nonzero components of the Riemann tensor:
\begin{equation}\label{fef-riem-rirj}
    \t{\hat{R}}{_\rho_i_\rho_j} = -\mri{S}_{ij} - 3\rho \mri{T}_{ij} + \order*{\rho^2},
\end{equation}
\begin{equation}\label{fef-riem-rijk}
    \t{\hat{R}}{_\rho_i_j_k} = -2\rho \mri{D}_{[j}\mri{S}_{k]i} + \order*{\rho^2},
\end{equation}
\begin{equation}
    \t{\hat{R}}{_i_j_k_l} = \t{\mri{R}}{_i_j_k_l} + \order*{\rho^2}.
\end{equation}
By virtue of (\ref{fef-riem-rirj}), the fact that $\mri{g}^{ij}\mri{T}_{ij} = 0$ \cite{kolanowski_hamiltonian_2021} and $\mri{S} = \mri{g}^{ij}\mri{S}_{ij} = \r{1}{4}\mri{R}$, we have
\begin{equation}
    \hat{R}_{\rho\rho} = \hat{g}^{ij}\hat{R}_{i\rho j\rho} = -\r{1}{4}\mri{R} + \order*{\rho^2}.
\end{equation}
Furthermore, from (\ref{fef-riem-rijk}) and the Bianchi identity $\mri{D}_{[i}\mri{R}_{jk]lm} = 0$ follows that
\begin{equation}
    \hat{R}_{\rho i} = \hat{g}^{jk}\hat{R}_{\rho j i k} = -\rho\f(\mri{D}_i \mri{S} - \mri{D}^j \mri{S}_{ji}) + \order*{\rho^2} = \order*{\rho^2}.
\end{equation}
Finally,
\begin{equation}
\begin{split}
    \hat{R}_{ij} &= \hat{g}^{\rho\rho}\hat{R}_{\rho i \rho j} + \hat{g}^{kl}\hat{R}_{kilj}
    = \mri{S}_{ij} + 3\rho\mri{T}_{ij} + \mri{R}_{ij} + \order*{\rho^2}
    = 2\mri{R}_{ij} - \r{1}{4}\mri{g}_{ij}\mri{R} + 3\rho\mri{T}_{ij} + \order*{\rho^2}
\end{split}
\end{equation}
and
\begin{equation}
    \hat{R} = \hat{g}^{\rho\rho}\hat{R}_{\rho\rho} + \hat{g}^{ij}\hat{R}_{ij}
    = \r{3}{2}\mri{R} + \order*{\rho^2}.
\end{equation}

Next, we calculate the Weyl tensor using
\begin{equation}
    \hat{C}_{\alpha\beta\gamma\delta}
    = \hat{R}_{\alpha\beta\gamma\delta}
    - \f(\hat{g}_{\alpha[\gamma}\hat{R}_{\delta]\beta} - \hat{g}_{\beta [\gamma}\hat{R}_{\delta]\alpha})
    + \r{1}{3}\hat{R}\hat{g}_{\alpha [\gamma}\hat{g}_{\delta]\beta},
\end{equation}
\begin{equation}\label{fef-weyl-rirj}
\begin{split}
    \hat{C}_{\rho i \rho j}
    &= \hat{R}_{\rho i \rho j} + \r{1}{2}\hat{R}_{ij} - \r{1}{2}\hat{g}_{ij}\hat{R}_{\rho\rho} - \r{1}{6}\hat{R}\hat{g}_{ij} \\
    &= -\mri{S}_{ij} - 3\rho\mri{T}_{ij} + \mri{R}_{ij} - \r{1}{8}\mri{R}\mri{g}_{ij} + \r{3}{2}\rho\mri{T}_{ij} + \r{1}{8}\mri{R}\mri{g}_{ij} - \r{1}{4}\mri{R}\mri{g}_{ij} + \order*{\rho^2} \\
    &= -\r{3}{2}\rho\mri{T}_{ij} + \order*{\rho^2},
\end{split}
\end{equation}
\begin{equation}\label{fef-weyl-rijk}
    \hat{C}_{\rho ijk} = \hat{R}_{\rho ijk} + \hat{g}_{i[j}\hat{R}_{k]\rho}
    = -2\rho \mri{D}_{[j}\mri{S}_{k]i} + \order*{\rho^2},
\end{equation}
\begin{equation}\label{fef-weyl-ijkl}
\begin{split}
    \hat{C}_{ijkl} &= \hat{R}_{ijkl} - \f(\hat{g}_{i[k}\hat{R}_{l]j} - \hat{g}_{j[k}\hat{R}_{l]i}) + \r{1}{3}\hat{R}\hat{g}_{i[k}\hat{g}_{l]j} \\
    &= \mri{R}_{ijkl} - 2\f(\mri{g}_{i[k}\mri{R}_{l]j} - \mri{g}_{j[k}\mri{R}_{l]i}) + \mri{R}\mri{g}_{i[k}\mri{g}_{l]j} - 3\rho\f(\mri{g}_{i[k}\mri{T}_{l]j} - \mri{g}_{j[k}\mri{T}_{l]i}) + \order*{\rho^2} \\
    &= \mri{C}_{ijkl} - 3\rho\f(\mri{g}_{i[k}\mri{T}_{l]j} - \mri{g}_{j[k}\mri{T}_{l]i}) + \order*{\rho^2} \\
    &= -3\rho\f(\mri{g}_{i[k}\mri{T}_{l]j} - \mri{g}_{j[k}\mri{T}_{l]i}) + \order*{\rho^2},
\end{split}
\end{equation}
where we used the fact that the Weyl tensor vanishes in three dimensions.

\section{\label{sec-bs}ASYMPTOTICALLY FLAT SPACETIMES IN BONDI-SACHS GAUGE}

One of the ways of describing an asymptotically flat spacetime in the neighborhood of $\scri$ (where $\scri$ is either the future or past null infinity) is using the Bondi-Sachs coordinates, where the physical metric satisfies \cite{sachs_gravitational_1962, madler_bondisachs_2016, strominger_lectures_2018}
\begin{equation}
\begin{gathered}
    g_{rr} = g_{rA} = 0, \quad
    g_{uu} = -1 + \r{2M}{r} + \order*{r^{-2}}, \quad
    g_{AB} = r^2\gamma_{AB} + rC_{AB} + \order*{1}, \\
    g_{ur} = -1 + \r{1}{16 r^2}C^{AB}C_{AB} + \order*{r^{-3}}, \quad
    g_{uA} = \r{1}{2}D^B C_{BA} + \order*{r^{-1}},
\end{gathered}
\end{equation}
where $M$ is the Bondi mass aspect and $C_{AB}$ contains the information about radiation at $\scri$ (it is equivalent to the asymptotic shear $\sigma^\circ$ in the notation of Ashtekar). The radial coordinate $r$ satisfies the condition
\begin{equation}
    \det\f(g_{AB}) = r^4\det\f(\gamma_{AB}),
\end{equation}
from which follows
\begin{equation}
    \gamma^{AB}C_{AB} = 0.
\end{equation}
Moreover, the Bondi news tensor is
\begin{equation}
    N_{AB} \coloneqq \pd_u C_{AB}.
\end{equation}

This formalism is equivalent to the method of conformal completion \cite{penrose_asymptotic_1963, geroch_asymptotic_1977, tamburino_gravitational_1966, tafel_comparison_2000}. After attaching the null boundary, one can introduce a coordinate system $(u,\Omega,x^A)$ on a neighborhood of $\scri$ which is related to the Bondi-Sachs by keeping the functions $u$ and $x^A$ the same and taking $\Omega = \r{1}{r}$. In those coordinates, we have
\begin{equation}\label{bs-metric-omega}
\begin{gathered}
    g_{\Omega\Omega} = g_{\Omega A} = 0, \quad
    g_{uu} = -1 + 2M\Omega + \order*{\Omega^2}, \quad
    g_{AB} = \Omega^{-2}\gamma_{AB} + \Omega^{-1}C_{AB} + \order*{1}, \\
    g_{u\Omega} = \Omega^{-2} - \r{1}{16}C^{AB}C_{AB} + \order*{\Omega}, \quad
    g_{uA} = \r{1}{2}D^B C_{BA} + \order*{\Omega},
\end{gathered}
\end{equation}
so the conformally rescaled metric $\hat{g} = \Omega^2 g$ extends to $\scri$. The inverse metric components satisfy
\begin{equation}\label{bs-metric-omega-inv}
\begin{gathered}
    g^{uu} = g^{uA} = 0, \quad
    g^{\Omega\Omega} = \Omega^4 - 2M\Omega^5 + \order*{\Omega^6}, \quad
    g^{AB} = \Omega^{2}\gamma^{AB} - \Omega^3 C^{AB} + \order*{\Omega^4}, \\
    g^{u\Omega} = \Omega^{2} + \r{1}{16}C^{AB}C_{AB}\Omega^4 + \order*{\Omega^5}, \quad
    g^{\Omega A} = -\r{1}{2}D_B C^{BA}\Omega^4 + \order*{\Omega^5}.
\end{gathered}
\end{equation}

From (\ref{bs-metric-omega}) and (\ref{bs-metric-omega-inv}) we calculate of the Christoffel symbols of $g$:
\begin{equation}
    \t{\Gamma}{^u_u_\Omega} = \t{\Gamma}{^u_\Omega_\Omega} = \t{\Gamma}{^u_\Omega_A} = 0,
\end{equation}
\begin{equation}
    \t{\Gamma}{^u_u_u} = \r{1}{2}g^{u\Omega}\f(2\pd_u g_{u\Omega} - \pd_\Omega g_{uu})
    = \r{1}{2}\Omega^2\f(-\r{1}{4}C^{AB}N_{AB} - 2M) + \order*{\Omega^3},
\end{equation}
\begin{equation}
    \t{\Gamma}{^u_u_A} = \r{1}{2}g^{u\Omega}\f(\pd_A g_{u\Omega} - \pd_\Omega g_{uA})
    = \order*{\Omega^2},
\end{equation}
\begin{equation}
    \t{\Gamma}{^u_A_B} = -\r{1}{2}g^{u\Omega}\pd_\Omega g_{AB}
    = \order*{\Omega^{-1}},
\end{equation}
\begin{equation}
    \t{\Gamma}{^\Omega_u_u} = \r{1}{2}g^{\Omega u}\pd_u g_{uu}
    + \r{1}{2}g^{\Omega\Omega}\f(2\pd_u g_{u\Omega} - \pd_\Omega g_{uu})
    + \r{1}{2}g^{\Omega B}\f(2\pd_u g_{uB} - \pd_B g_{uu})
    = \order*{\Omega^3},
\end{equation}
\begin{equation}
    \t{\Gamma}{^\Omega_u_A} = \r{1}{2}g^{\Omega u}\pd_A g_{uu}
    + \r{1}{2}g^{\Omega\Omega}\f(\pd_A g_{u\Omega} - \pd_\Omega g_{uA})
    + \r{1}{2}g^{\Omega B}\f(\pd_u g_{AB} + 2\pd_{[A} g_{B]u})
    = \order*{\Omega^3},
\end{equation}
\begin{equation}
    \t{\Gamma}{^A_u_u} = \r{1}{2}g^{A\Omega}\f(2\pd_u g_{u\Omega} - \pd_\Omega g_{uu})
    = \order*{\Omega^4},
\end{equation}
\begin{equation}
    \t{\Gamma}{^A_u_\Omega} = \r{1}{2}g^{AB}\f(\pd_\Omega g_{uB} - \pd_B g_{u\Omega})
    = \order*{\Omega^2},
\end{equation}
\begin{equation}
    \t{\Gamma}{^A_u_B} = \r{1}{2}g^{AC}\f(\pd_u g_{BC} + 2\pd_{[B} g_{C]u})
    = \r{1}{2}\t{N}{^A_B}\Omega + \order*{\Omega^2},
\end{equation}
\begin{equation}
    \t{\Gamma}{^A_\Omega_B} = \r{1}{2}g^{AC}\pd_\Omega g_{BC} = \order*{\Omega^{-1}},
\end{equation}
\begin{equation}
    \t{\Gamma}{^A_B_C} = \r{1}{2}g^{AD}\f(2\pd_{(B}g_{C)D} - \pd_D g_{BC}) - \r{1}{2}g^{A\Omega}\pd_\Omega g_{BC} = \order*{1}.
\end{equation}
Next, we derive some components of the Riemann tensor of $g_{ab}$:
\begin{equation}
    \t{R}{^u_u_u_\Omega} = 2\pd_{[u}\t{\Gamma}{^u_{\Omega]}_u} + 2\t{\Gamma}{^u_e_{[u}}\t{\Gamma}{^e_{\Omega]}_u} = \Omega\f(\r{1}{4}C^{AB}N_{AB} + 2M) + \order*{\Omega^2},
\end{equation}
\begin{equation}
    \t{R}{^u_u_u_A} = 2\pd_{[u}\t{\Gamma}{^u_{A]}_u} + 2\t{\Gamma}{^u_e_{[u}}\t{\Gamma}{^e_{A]}_u}
    = \order*{\Omega^2},
\end{equation}
\begin{equation}
    \t{R}{^A_u_u_B} = 2\pd_{[u}\t{\Gamma}{^A_{B]}_u} + 2\t{\Gamma}{^A_e_{[u}}\t{\Gamma}{^e_{B]}_u}
    = \order*{\Omega^2} = \r{1}{2}\Omega\pd_u \t{N}{^A_B}.
\end{equation}
Since $R_{ab} = 0$, we have $\t{C}{^a_b_c_d} = \t{R}{^a_b_c_d}$. Therefore
\begin{equation}
    \t{C}{^u_u_u^u} = \t{\hat{C}}{^u_u_u_\Omega} + \order*{\Omega^2}
    = \Omega\f(\r{1}{4}C^{AB}N_{AB} + 2M) + \order*{\Omega^2},
\end{equation}
\begin{equation}
    \t{C}{^u_u_u^A} = \hat{g}^{AB}\t{\hat{C}}{^u_u_u_B}
    = \f(\gamma^{AB} + \order*{\Omega})\t{\hat{C}}{^u_u_u_B} + \order*{\Omega^2}
    = \order*{\Omega^2},
\end{equation}
\begin{equation}
    \t{C}{^A_u_u^B} = \hat{g}^{BC}\t{\hat{C}}{^A_u_u_C}
    = \f(\gamma^{BC} + \order*{\Omega})\t{\hat{C}}{^A_u_u_C} + \order*{\Omega^2}
    = \r{1}{2}\Omega\pd_u N^{AB} + \order*{\Omega^2}.
\end{equation}

\bibliography{main.bib}

\end{document}